\documentclass[%
 reprint,
superscriptaddress,
 amsmath,amssymb,
 aps,
]{revtex4-2}

\pdfoutput=1

\usepackage[utf8]{inputenc}
\usepackage[T1]{fontenc}

\usepackage{graphicx}
\usepackage{dcolumn}
\usepackage{bm}
\usepackage{color}
\usepackage{graphics}
\usepackage{subfigure}
\usepackage{pdftexcmds}
\usepackage{ifpdf}
\usepackage{braket}
\usepackage{dsfont}

\usepackage{url}

 \newcommand{\bc}{\begin{center}}
 \newcommand{\ec}{\end{center}}
 
\newcommand{\be}{\begin{equation}}
\newcommand{\ee}{\end{equation}}

\newcommand{\bea}{\begin{eqnarray}}
\newcommand{\eea}{\end{eqnarray}}

\newcommand{\bd}{\begin{displaymath}}
\newcommand{\ed}{\end{displaymath}}

\newcommand{\iu}{\mathrm{i}}

\newcommand{\hc}{h_\mathrm{c}}

\newcommand{\hn}{\hat{n}}

\newcommand{\hx}{\hat{x}}
\newcommand{\hy}{\hat{y}}
\newcommand{\hz}{\hat{z}}
\newcommand{\hs}{\hat{s}}
\newcommand{\he}{\hat{e}}

\newcommand{\Be}{\vec{B}_{\mathrm{eff}}}
\newcommand{\bef}{\vec{b}_{\mathrm{eff}}}

\newcommand{\atan}{\,\mathrm{atan}\,}

\newcommand{\asinh}{\,\mathrm{asinh}\,}
\newcommand{\atanh}{\,\mathrm{atanh}\,}
\newcommand{\sn}{\mathrm{sn}}
\newcommand{\cn}{\mathrm{cn}}
\newcommand{\dn}{\mathrm{dn}}
\newcommand{\cd}{\mathrm{cd}}

\newcommand{\e}{\text{e}}

\newcommand{\pmin}{p_\text{min}}
\newcommand{\pmax}{p_\text{max}}

\newcommand{\emod}{\eta}
\newcommand{\nome}{q}

\begin{document}

\title{Continuum of metastable helical states in monoaxial chiral magnets: effect of the boundary conditions}

\author{V. Laliena}
\affiliation{
Department of Applied Mathematics and
Institute of Mathematics and Applications (IUMA), University of Zaragoza
C/ Mar\'ia de Luna 3, 50018 Zaragoza, Spain
}

\author{S. A. Osorio}
\affiliation{ 
Instituto de Nanociencia y Nanotecnolog\'ia (CNEA-CONICET), Nodo Bariloche, Av. Bustillo 9500 (R8402AGP), S. C. de Bariloche, R\'io Negro, Argentina
}
\affiliation{
Gerencia de F\'isica,  Centro At\'omico Bariloche, Av. Bustillo 9500 (R8402AGP), S. C. de Bariloche, R\'io Negro, Argentina
}

\author{S. Bustingorry}
\affiliation{ 
Instituto de Nanociencia y Nanotecnolog\'ia (CNEA-CONICET), Nodo Bariloche, Av. Bustillo 9500 (R8402AGP), S. C. de Bariloche, R\'io Negro, Argentina
}
\affiliation{
Gerencia de F\'isica,  Centro At\'omico Bariloche, Av. Bustillo 9500 (R8402AGP), S. C. de Bariloche, R\'io Negro, Argentina
}
\affiliation{
Aragon Nanoscience and Materials Institute (CSIC-University of Zaragoza) and Condensed Matter Physics Department, University of Zaragoza, C/ Pedro Cerbuna 12, 50009 Zaragoza, Spain
}

\author{J. Campo}
\affiliation{
Aragon Nanoscience and Materials Institute (CSIC-University of Zaragoza) and Condensed Matter Physics Department, University of Zaragoza, C/ Pedro Cerbuna 12, 50009 Zaragoza, Spain
}
\affiliation{International Institute for Sustainability with Knotted and Chiral Meta Matter  (Visiting Professor), 2-313 Kagamiyama, Higashi-Hiroshima, Hiroshima, 739-0046, Japan}  

\date{\today}

\begin{abstract}
In a recent publication we showed that a monoaxial chiral magnet has a continuum of metastable helical states differing by the helix wave number. This intringuing result was obtained for the case of an infinite magnet (or of a magnet with periodic boundary conditions). However, it has been pointed out that in a real magnet only one of these states is compatible with the  boundary conditions, because the helix wave number is determined by the surface chiral twist. Thus, only one of the continuum of states is physically realizable. This is true for the case of a chiral magnet in contact with a non magnetic medium (vacuum or air, for instance), but the boundary conditions can be altered by setting the chiral magnet in contact with another magnetic medium, which may be able to absorb the surface chiral twist. We show here that this is indeed the case by studying a composite magnet system, which consists of one monoaxial chiral magnet of rectangular parallelepiped shape which has two similar uniaxial ferromagnets attached to each of the faces that are perpendicular to the chiral axis. We show that, in the case of zero applied field, this composite system has a number of metastable helical states which is proportional to the length $L_0$ of the chiral magnet along the chiral axis, and that the results of our previous publication are recovered in the limit $L_0\to\infty$.
\end{abstract}

\maketitle

\section{Introduction}

Chiral magnets, characterized by the presence of a sizable Dzyaloshinskii-Moriya interaction (DMI), are being extensively studied since they host noncollinear magnetic states which appear as metastable or equilibrium states at low temperature. Besides their intrinsic theoretical interest, these magnetic textures have inportant potential applications in spintronics and magnonics \cite{Back2020,Barman2021,Mruczkiewicz2021,Vedmedenko2020}. Examples of these noncollinear magnetic textures are the skyrmions of cubic chiral magnets \cite{Bogdanov1994b,Nagaosa2013}, the one dimensional chiral solitons of monoaxial chiral magnets \cite{Dzyal1964,Kishine2015}, and the helical or conical states which appear in both types of chiral magnets 
\cite{Bak1980,Kishine2015}.

Cubic chiral magnets have been much thoroughly studied than monoaxial chiral magnets, the object of the present work, but nevertheless the main features of the equilibrium properties of the latter are rather well understood.
In monoaxial chiral magnets the DMI acts only along one specific direction with coincides with one crystallographic axis. We call this direction the \textit{chiral axis}. 
To put this work in its context, let us summarize briefly the equilibrium properties of monoaxial chiral magnets. At low temperature and zero applied field the equilibrium state is a helical structure whose wave number is determined by the competition between the Heisenberg exchange interaction and the DMI. When an external field is applied the helical structure becomes a conical structure if the field is parallel to the chiral axis, a chiral soliton lattice if the field is perpendicular to the chiral axis, or a magnetic structure which interpolates between these two limiting cases if the field is neither perpendicular nor parallel to the chiral axis
\cite{Miyadai1983,Togawa2012,Laliena2016a,Laliena2016b,Laliena2017a,Ghimire2013,Chapman2014,
Tsuruta2016,Yonemura2017,Clements2017,Clements2018,Osorio2023}. If the applied field strength is high enough the equilibium state is a forced ferromagnetic state, which can host metastable isolated solitons \cite{Laliena2020}.

Although in the last years the magnetic states that have received more attention and caused more excitement have been the topologically non trivial skyrmions, recently there has been a revival of the interest in the helical/conical states, since the conical phases occupy a larger fraction of the phase diagram and thus are created more easily. In cubic chiral magnets the equilibrium helical states are highly degenerated since the helix wave vector can point in any direction. It has been shown that the wave vector direction can be controlled by electric means and thus helical states with different wave vectors can be used to storage and manipulate information  \cite{Masell2020prb}. The direction of the helix wave vector can also be changed by means of thermal currents \cite{Yasin2023}. Therefore, the orientation of helical stripes may serve as building blocks for devices for classical or unconventional computing, in what would be a new technology that may be named \textit{helitronics} \cite{Bechler2023}.

In monoaxial chiral magnets the degeneracy of the helical state is absent since the helix wave vector points along the chiral axis in the direction determined by the DMI. However, it was shown that in monoaxial chiral magnets there is a continuum of metastable helical states differing by the helix wave number \cite{Laliena2023,Laliena2018a} (metastable helical states of this kind exists also in cubic chiral magnets \cite{Laliena2017b}). In reference \onlinecite{Laliena2023} we showed that it is possible to swicth between these helical states, which were called there $p$ states, by applying magnetic fields and electric currents along the chiral axis. Therefore, these $p$ states may serve as building blocks for computing devices.

The existence of this continuum of metastable states, differing from the equilibrium helical state only by the wave number, is an intringuing question. Theses states were obtained by solving the magnetic equilibrium equations for an infinite magnet, ignoring thus the boundary conditions. However, it has been pointed out that in a real magnet the surface chiral twist required by the boundary conditions in presence of DMI actually selects the equilibrium helical state, which is the only $p$ state which satisfies the boundary conditions \cite{Garst_private}. This is true for a chiral magnet which is surrounded by a non magnetic medium (air or vacuum, for instance). But if the chiral magnet is in contact with another magnet with appropriate characteristics, the surface chiral twist induced by the DMI can be absorbed by the surrounding magnet, and the continuum of metastable $p$ states is present. The purpose of this work is to show that this idea is indeed realized. For simplicity, we restrict the analysis to the case of zero applied field and zero current, because in this case the problem can be solved exactly. The general case, with non zero applied field and/or current, requires a numerical treatment and will be addressed later.

The paper is organized as follows. In section \ref{sec:bc} we analize carefully the conditions that the magnetization has to satisfy at the interface that separates two different media; in section \ref{sec:model} we describe a system which host a continuum of $p$ states, which consists of a monoaxial chiral magnet of rectangular shape, with the two faces perpendicular to the chiral axis sticked to two similar uniaxial ferromagnets; section \ref{sec:p} is devoted to the determination of the helical states of this system and in section \ref{sec:stab} the stability of these states is analyzed. Finally, in section \ref{sec:conc} we summarize the conclusions.

\section{Conditions at the interface between the two media \label{sec:bc}}

The magnetization of a composite magnetic system formed by several magnetic media set in contact is not a smooth function in general, but it is generically discontinuous at the interfaces, due to the discontinuity of the saturation magnetization. However, the mathematical structure of the Landau-Lifschitz-Gilbert (LLG) equation set constraints on the nature of the singularity of the magnetization. It turns out that the vector field that describe the direction of the magnetization has to be continuous at the interface, although its derivative along the normal vector of the interface may be discontinuous. 

The conditions which has to satisfy the magnetization at interfaces have been obtained in reference \onlinecite{Abert2019} (see also reference \onlinecite{Heistracher2022}) by studying the variational problem from which the LLG equation is derived.
In this section we analyze these conditions directly from the LLG equation, rather than from the variational approach. 

To set the notation, let the unit vectors $\hx=\hx_1$, $\hy=\hx_2$ and $\hz=\hx_3$ form a set of cartesian coordinate axes and let $x=x_1$, $y=x_2$, and $z=x_3$ be the corresponding coordinates.
For notational convenience, to analyze the conditions on the magnetization at the interface between two different media it is convenient to work with a system which is slightly more general than that studied in this work (section \ref{sec:model}). 
Thus, in this section we consider an inhomogeneous magnet in which the magnetization direction is described by the unit vector field $\hn$ and whose energy density is given by
\be
W = \sum_{i=1}^3\Big(A\partial_i\hn\cdot\partial_i\hn - D_i\hn\cdot(\hx_i\times\partial_i\hn) \Big)+ W_0(\hn),
\ee
where $W_0$ contains terms which do not depend on the derivatives of $\hn$ and $A$ and $D_i$ are the intensities of the ferromagnetic and DMI exchange interactions, respectively. If $D_1=D_2=D_3$ we have a chiral cubic magnet and if $D_1=D_2=0$ and $D_3\neq0$ we have a monoaxial chiral magnet with chiral axis along $\hz$. 
The interaction intensities $A$ and $D_i$ are smooth functions of the position and depend on a parameter $\delta>0$ in such a way that in the limit $\delta\to0$ they become sharp (discontinuous) at some given surface.
So, for small $\delta$ the system consists of two different media in contact, separated by the mentioned surface.

The effective field, obtained from the functional derivative of the energy with respect to $\hn$, can be written as $\Be=\Be^{(d)}+\Be^{(0)}$, where $\Be^{(0)}$ does not depend on the derivatives of $\hn$ and
\be
\Be^{(d)}= \frac{1}{M_s}\sum_i\Big(
\partial_i(2A\partial_i\hn) - D_i\hx_i\times\partial_i\hn - \hx_i\times\partial_i(D_i\hn)\Big).
\ee
In the above equation
$M_s$ is the saturation magnetization, which is a smooth function of the position which may become discontinuous at the interface for $\delta\to0$.

In the limit of a sharp interface ($\delta\to0$) the magnetization may become a non smooth function which, nevertheless, has to fullfill some conditions which are derived from the structure of the Landau-Lifschitz-Gilbert (LLG) equation, which has the form
\be
\partial_t\hn = \gamma\Be\times\hn + \alpha\hn\times\partial_t\hn,
\ee
where $\gamma>0$ is the absolute value of the electron giromagnetic ratio and $\alpha$ is the Gilbert damping parameter. Using the product rule for derivatives, the term $\Be^{(d)}\times\hn$ can be written as
\begin{gather}
\Be^{(d)}\times\hn = \frac{1}{M_s}\sum_i\left[\partial_i\Big((2A\partial_i\hn-D_i\hx_i\times\hn)\times\hn\Big)\right. \nonumber \\[4pt]
+ \left. D_i\Big((\hx_i\times\hn)\times\partial_i\hn - (\hx_i\times\partial_i\hn)\times\hn\Big)\right].
\label{eq:intparts}
\end{gather}

\begin{figure}[t!]
\includegraphics[width=0.3\textwidth]{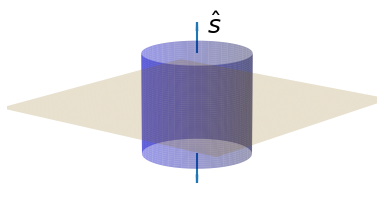}
\caption{A small cylindric pillbox enclosing a surface element of the interface, with the axis oriented along the normal of the surface element, $\hat{s}$.
\label{fig:pillbox}}
\end{figure}

To find the conditions at the interface, we consider a small cylindric pillbox which encloses a surface element of the interface, and whose axis is oriented along the normal vector of the surface element, $\hs$, as in Fig.~\ref{fig:pillbox}. The pillbox occupies the volume $\Omega_p$ and is bounded by the surface $\partial\Omega_p$. We multiply both sides of the  LLG equation by $M_s$ and integrate over $\Omega_p$. Using the divergence theorem for the first term of the right hand side of equation (\ref{eq:intparts}), we get
\begin{gather}
\int_{\Omega_p}\!\!\! M_s\partial_t\hn\,d^3 r = \gamma\!\int_{\partial\Omega_p} \!\!\!d\sigma \sum_i \sigma_i(2A\partial_i\hn-D_i\hx_i\times\hn)\times\hn \nonumber \\[4pt]
+\int_{\Omega_p}\!\!\gamma\left[
\sum_i D_i \Big((\hx_i\times\hn)\times\partial_i\hn - (\hx_i\times\partial_i\hn)\times\hn\Big)
\right] d^3r \nonumber \\[4pt]
+\int_{\Omega_p} \!M_s\left[\gamma\Be^{(0)}\times\hn + \alpha\hn\times\partial_t\hn 
\right]d^3 r, \label{eq:pillbox}
\end{gather}
 where $\hat\sigma$ is the normal vector of $\partial\Omega_p$. Now, we take the limit of sharp interface, $\delta\to0$, making the following assumptions: 1) $\hn$ remains continuous at all points, including the interface; 2) the derivatives $\partial_i\hn$, $\partial_t\hn$ remain bounded, although $\partial_i\hn$ may be discontinuous at the interface. Next, we take the limit in which the pillbox thinkness tends to zero. In this limit the volume integrals and the surface integral over the curved face of the pillbox vanish, and thus equation (\ref{eq:pillbox}) requires that
 \be
 \vec{C} = 2A(\hs\cdot\nabla)\hn-(\tilde{D}\hs)\times\hn 
 \ee
 be continuous at the interface. Here $\tilde{D}$ is the diagonal $3\times 3$ matrix with $D_1$, $D_2$, and $D_3$ in the diagonal. Continuity, obviously, means that
 \be
 \lim_{t\to0^+} \vec{C}(\vec{x}-t\hat{s})  = \lim_{t\to0^+} \vec{C}(\vec{x}+t\hat{s}) \label{eq:cond_gen}
 \ee
for any $\vec{x}$ at the interface. A consequence of this fact is that if $A$ or $D$ are discontinuous at the interface, then the derivative of $\hn$ along the surface normal has to be discontinuous at the interface.
 
For a cubic chiral magnet $D_1=D_2=D_3=D$ and then
 \be
 \vec{C} = 2A(\hs\cdot\nabla)\hn- D\hs\times\hn, \label{eq:cond_cubic}
 \ee
while for a monoaxial chiral magnet with chiral axis along $\hz$ we have $D_1=D_2=0$, $D_3=D$, and then
 \be
  \vec{C} =2A(\hs\cdot\nabla)\hn- D(\hz\cdot\hs)\hz\times\hn. \label{eq:cond_mono}
 \ee

If the magnet is in contact with a non magnetic medium (vacuum or air, for instance), then $\vec{C}$ has to vanish at the boundary, because it vanishes in the non magnetic medium, since there $A$ and $D$ vanish. This is the boundary condition which originates the well known surface chiral twists.
 
 On the other hand, if a magnet (chiral or not) is in contact with a very hard magnet, the expression (\ref{eq:cond_gen}) has to be equated to $A_h(\hs\cdot\nabla)\hn$, which corresponds to the hard magnet side. If the stiffness constant of the hard magnet, $A_h$, is very large, then $(\hs\cdot\nabla)\hn$ has to be proportionally small on the hard magnet side, and it vanishes in the limit $A_h\to\infty$.   In this limit $A_h(\hs\cdot\nabla)\hn$ take the value required by the right hand side so that $\hn$ has  direction of the equilibrium magnetization of the hard magnet. We obtain in this way Dirichlet boundary conditions.
 
As a word of caution, let us notice that the discussion on conditions at interfaces presented here, including boundary conditions, ignores the possible existence of surface anisotropies, in which case the condition at the interface would be
\be
\lim_{t\to0^+} \Big(\vec{C}(\vec{x}+t\hat{s}) - \vec{C}(\vec{x}-t\hat{s})\Big) = \vec{S},
\label{eq:cond_surfanis}
\ee
where $\vec{S}$ is the contribution of the surface anisotropy to the volume integrals of (\ref{eq:pillbox}), which does not vanish in the limit of sharp interface ($\delta\to0$) and infinitely thin pillbox, in this order. Hence, $\vec{C}$ could be discontinuous at the interface.

Finally, it is worthwhile to stress that at a sharp interface that separates two magnetic media the saturation magnetization becomes discontinuous, what induces a surface density of magnetic charge at each point $\vec{x}$ of the surface, given by 
\be
\lim_{t\to0^+}\Big(M_s(\vec{x}-t\hs)-M_s(\vec{x}+t\hs)\Big)\hs\cdot\hn.
\ee
This surface magnetic charge contributes to the magnetostatic field of the two media, but does not affect the interface conditions given by (\ref{eq:cond_gen}) or (\ref{eq:cond_surfanis}) \cite{Abert2019}.

\section{A composite magnetic system \label{sec:model}}

\begin{figure}[t!]
\includegraphics[width=0.3\textwidth]{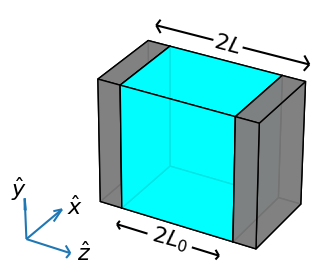}
\caption{Composite magnet. The  cyan region is occupied by a monoaxial chiral magnet and the gray regions by two similar uniaxial ferromagnets. The chiral axis is oriented along the $\hz$ direction.
\label{fig:composite_magnet}}
\end{figure}

In this work, we consider a magnetic system of rectangular parallelepiped shape which occupies a region of size $2L$ along the $\hz$ direction, so that $-L\leq z\leq L$ (the convention for the coordinate system is described at the beginning of section \ref{sec:bc}). The dimensions of the system in the directions $\hx$ and $\hy$ are very large and thus are considered infinite. The system is inhomogeneous along the $\hz$ direction and consists of three homogeneous parts: one monoaxial chiral magnet occupies a central region of size $2L_0$, that is, the region $-L_0\leq z\leq L_0$; the peripheral regions, $-L\leq z < -L_0$ and $L_0 < z \leq L$,  are occupied by two similar uniaxial ferromagnets,  as in figure \ref{fig:composite_magnet}. The monoaxial chiral magnet is oriented so that its chiral axis is aligned with the $\hz$ axis, and the easy axis of each uniaxial ferromagnet is aligned with the $\hx$ axis. The direction of the magnetization is described by the unit vector field $\hn$.

It is convenient to introduce the characteristic functions $\chi_c(z)$ and $\chi_u(z)$ such that
$\chi_c(z)=1$ if $|z|\leq L_0$ and $\chi_c(z)=0$ otherwise, and $\chi_u(z)=1$ if $L_0<|z|\leq L$ and $\chi_u(z)=0$ otherwise. The energy of the system is given by 
\be
E=\int d^3r \, (\chi_cW_c+\chi_uW_u),
\ee
where $W_c$ and $W_u$ are the energy densities of the monoaxial chiral magnet and of the uniaxial ferromagnet, respectively, and are given by
\begin{gather}
\hspace*{-0.2cm}W_c =A \sum_i\partial_i\hn\cdot\partial_i\hn - D\hz\cdot(\hn\times\partial_z\hn) - K_c(\hz\cdot\hn)^2, \\[2pt]
\hspace*{-0.5cm}W_u = \rho A \sum_i\partial_i\hn\cdot\partial_i\hn - K_u(\hx\cdot\hn)^2.
\end{gather}
In the above expressions $A$ is the stiffeness constant of the chiral magnet and the dimensionless parameter $\rho$ is the ratio beween the stiffness constants of the uniaxial ferromagnet and the chiral magnet. 
The chiral magnet has a uniaxial anisotropy, which is of easy-plane type, whose axis coincides with the chiral axis $\hz$, and whose energy per unit volume is given by the anisotropy constant $K_c<0$. The uniaxial magnet has its easy axis along  $\hx$ and its anisotropy constant is $K_u>0$. 
Finally, $D$ sets the strength of the DMI interaction in the chiral magnet. 
We ignore the magnetostatic energy since it can be included in the anisotropies for the one dimensional modulations considered in this work \cite{Hubert2008}. Notice also that we consider only the case of zero applied field.

The effective field is given by $\Be=-(1/M_s)\delta E/\delta\hn$, where the saturation magnetization $M_s$ is a function of $z$ given by
$M_s(z) = M_c \chi_c(z) + M_u\chi_u(z)$,
where $M_c$ and $M_u$ are the saturation magnetizations of the chiral magnet and the uniaxial ferromagnet, respectively. If $\hn$ satisfies the conditions discussed in section \ref{sec:bc}, that is, continuity of $\hn$ and $\vec{C}$, integration by parts can be applied to obtain the functional derivative in the standard way, and we obtain $\Be=(2A/M_s)\bef$, where
\be
\begin{split}
\hspace*{-0.25cm}
\bef = a\nabla_T^2 & \hn + \partial_z \big(a\partial_z\hn  - q_0\chi_c\hz\times\hn\big)
- q_0\chi_c\hz\times\partial_z\hn   \\[2pt]
&+ q_0^2\kappa\chi_c(\hz\cdot\hn)\hz + \rho q_u^2\chi_u(\hx\cdot\hn)\hx.
\label{eq:beff}
\end{split}
\ee
In the above expresion we introduced $\nabla_T^2=\partial_x^2+\partial_y^2$, 
\be
q^2_0 = \frac{D}{2A}, \quad \kappa=\frac{AK_c}{D^2}, \quad q_u^2=\frac{K_u}{\rho A},
\ee
and  the function $a(z) = \chi_c(z) + \rho\chi_u(z)$.

At the interfaces $z=\pm L_0$, both $\hn$ and $\vec{C}$, given by (\ref{eq:cond_mono}) with $\hs=\hz$, have to be continuous. The condition of continuity of $\vec{C}$ at $z=L_0$ can be cast to the form
\be
\lim_{z\to L_0^-} \big(\partial_z\hn-q_0\hz\times\hn\big) = \lim_{z\to L_0^+} \rho\,\partial_z\hn.
\label{eq:matching}
\ee
An analogous condition holds for $z=-L_0$. We call these conditions at $z=\pm L_0$ the \textit{matching conditions}. 

Finally, the uniaxial ferromagnets are in contact at $\pm L$ with a non magnetic medium, what means that expression (\ref{eq:cond_gen}), with $\hs=\hz$ and $\tilde{D}=0$, has to vanish at the boundaries $z=\pm L$. This provides the Neumann boundary conditions
\be
\partial_z\hn(-L) = \partial_z\hn(L) = 0. \label{eq:bc}
\ee
 
\section{Helical states \label{sec:p}}

Since we seak static magnetic states with modulations only along the $\hz$ direction, the effective field 
can be written as
\be
\bef = \left\{
\begin{array}{l}
\hn^{\prime\prime} - 2q_0\hz\times\hn^\prime + q_0^2\kappa(\hz\cdot\hn)\hz, \;
|z|<L_0, \\[4pt]
\rho\big(\hn^{\prime\prime} + q_u^2(\hx\cdot\hn)\hx\big), \;
L_0<|z|<L,
\end{array}
\right.
\label{eq:beffsplit}
\ee
where the prime means derivative with respect to $z$. 

The equation for the static states is $\bef\times\hn=0$. 
Equation (\ref{eq:beffsplit}) implies that we have to solve one differential equation for $|z|<L_0$ and another one for $L_0<|z|<L$, and to impose the matching condition (\ref{eq:matching}) at $z=L_0$ and the analogous condition for $z=-L_0$, and the boundary conditions (\ref{eq:bc}).

The static equations admit solutions in which the magnetization lies on the easy plane of the chiral magnet, and thus can be written as
\be
\hn = \cos\varphi\,\hx + \sin\varphi\,\hy, \label{eq:hn}
\ee
where the function $\varphi(z)$ is a solution of
\begin{gather}
\varphi^{\prime\prime} = 0, \quad |z|<L_0, \label{eq:helixde} \\[0pt]
\varphi^{\prime\prime} - q_u^2\sin\varphi\cos\varphi = 0, \quad L_0<|z|<L, \label{eq:sg}
\end{gather}
which satisfies the boundary conditions and the matching conditions.

The general solution of (\ref{eq:helixde}) is $\varphi(z)=\varphi(0)+pq_0z$, where $\varphi(0)$ and $p$ are arbitrary constants, while equation (\ref{eq:sg}) is the well known double Sine-Gordon equation, which also has a known two parameter family of solutions. This allows us to construct exact solutions for the whole system, which satisfy the differential equations (\ref{eq:helixde}) and  (\ref{eq:sg}), the boundary conditions, and the matching conditions. Specifically, we propose a symmetric solution which has helical nature within the chiral magnet, given by
\be
\varphi(z) = \left\{
\begin{array}{cc}
-\sigma_p\,\varphi_0(-z), & \;-L<z<-L_0, \\[4pt]
pq_0z, & \;-L_0<z<L_0, \\[4pt]
\sigma_p\,\varphi_0(z), & \;L_o<z<L, 
\end{array}
\right.
\label{eq:phi}
\ee
where the parameter $p$, which is the helix wave number in units of $q_0$, is to be determined. In equation (\ref{eq:phi}) we introduce $\sigma_p=1$ if $p\geq1$ and $\sigma_p=-1$ if $p<1$. As it will become clear in the  following, $\sigma_p$ is needed to satisfy the matching conditions (\ref{eq:matching}).
The function $\varphi_0(z)$ is the solution of equation (\ref{eq:sg}) with $-\pi<\varphi_0<0$ and
\be
\varphi_0(z_0)=-\frac{\pi}{2}, \quad \varphi_0^\prime(L) = 0, \quad \varphi_0^\prime(z)>0, \label{eq:condphi0}
\ee
where $z_0<L$ is a point to be determined by imposing the matching conditions. 
The explicit form of $\varphi_0$ is obtained in appendix \ref{app:sg}, and is given by
\be
\varphi_0(z) = -\arccos\Big(\emod\,\sn\big(q_u(z-z_0),\emod\big)\Big), \label{eq:phi0}
\ee
where $\sn(x,\emod)$ is the Jacobi elliptic function, with ellipticity modulus $\emod$, and $z_0<L$ is chosen such that $\varphi_0(z_0)=-\pi/2$. If $L-L_0$ is large, we may visualize $\varphi_0$ as a domain wall centered at $z_0$, which connects two domains with magnetization pointing along $\pm\hx$ for $z\to\pm\infty$.

Equation (\ref{eq:phi0}) is complemented with
\be
K(\emod) = q_u(L-z_0), \label{eq:K}
\ee
where $K(\emod)$ is the complete elliptic integral of the first kind. The above equation, which determines the parameter $\emod$, ensures that the boundary condition $\varphi^\prime(L)=0$ is satisfied (see appendix \ref{app:sg}). Taking into account that $\sn\big(K(\emod),\emod\big)=1$, equation (\ref{eq:phi0}) gives $\emod=\cos\varphi_0(L)$.

Summarizing, the magnetic state given by (\ref{eq:phi}) consists of a helical state of wave number $pq_0$ within the  chiral magnet connected at $z=L_0$ to a section of a domain wall hosted by the uniaxial ferromagnet in the $z>L_0$ region. The wall center, $z_0$, is a free parameter tuned to enforce the matching conditions. The helical state is also connected to a section of another domain wall section hosted by the uniaxial ferromagnet in the $z<-L_0$ region. This latter domain wall is obtained from the former domain wall by a symmetry. The domain wall center, $z_0$, need not be at a physical point inside the uniaxial magnet, but can lie in the region $z<L_0$. Actually, this view of the magnetic states in the uniaxial ferromagnets as sections of domain walls holds only if these magnets are thick enough. However, we find it useful to think of these magnetic states as domain walls.

The continuity of $\hn(z)$ is guaranteed if and only if
\begin{gather}
\cos\varphi_0(L_0) = \cos(pq_0L_0), \label{eq:cont_cos} \\[2pt]
\sigma_p\sin\varphi_0(L_0) = \sin(pq_0L_0), \label{eq:cont_sin}
\end{gather}
that is
\be
\sigma_p \varphi_0(L_0) = (pq_0L_0)\,\,\text{mod}\, 2\pi. \label{eq:contphi}
\ee
The matching condition (\ref{eq:matching}) reduces to
\be
\rho \sigma_p \varphi_0^\prime(L_0) = (p-1)q_0, \label{eq:pphip}
\ee
which, taking into account the form of $\varphi_0^\prime$ (appendix \ref{app:sg}), the definition of $\sigma_p$, and equation (\ref{eq:cont_cos}), can be cast to the form
\be
|p-1| = \frac{\rho q_u}{q_0}\sqrt{\emod^2-\cos^2(pq_0L_0)}. \label{eq:p}
\ee
By symmetry, the matching condition at $z=-L_0$ is also satisfied if equation (\ref{eq:p}) holds.

Equations (\ref{eq:K}), (\ref{eq:contphi}), and (\ref{eq:p}) determine completely the magnetic states of the form (\ref{eq:phi}). They constitute a system of three equations with three unknowns: $\emod$, $z_0$, and $p$. Since $0\leq\emod<1$, the right hand side of (\ref{eq:p}) is bounded by $\rho q_u/q_0$, what implies the following bounds for $p$:
\be
1 - \frac{\rho q_u}{q_0} \leq p \leq 1 + \frac{\rho q_u}{q_0}. \label{eq:bounds}
\ee

We argue below that equations (\ref{eq:K}), (\ref{eq:contphi}), and (\ref{eq:p}) have many solutions, with different values of $p$, if $L_0$ and $L-L_0$ are large. The numerical solution of those equations indicates that this is true also if $L-L_0$ is not large. Therefore, there are many states of the form (\ref{eq:phi}) differing by the wave number $pq_0$ of the helical part (the magnetization of the uniaxial ferromagnets is also different for different values of $p$, of course). The values of $p$ for these states become dense in a certain interval in the limit $L_0\to\infty$ (however $L-L_0$ may remain finite).

To work in the large $L-L_0$ limit, which simplifies the analysis considerably, it is convenient to substitute the ellipticity modulus $\emod$ by the \textit{nome}, $\nome=\exp\big(-\pi K/\bar{K}\big)$, where, 
\be
K=K(\eta), \quad \bar{K} = K\big(\sqrt{1-\emod^2}\big). \label{eq:KKb}
\ee
Notice that, in the realm of elliptic functions, the conventional notation for what we call $\bar{K}$ is $K^\prime$. We depart from the conventional notation to avoid confusion, since we use the prime for derivatives. 
Hence, from now on we consider that $\eta$ is a function of $q$, given by inverting the equation that defines the nome. Notice that equation (\ref{eq:K}) means that the nome is exponentially small, $q\sim\exp\big(-2q_u(L-L_0)\big)$, for large $L-L_0$.
Using the properties of the complete elliptic integral $K$ \cite{Gradshteyn2007}, we see that $\emod = 1 + O(q)$ and then equation (\ref{eq:p}) has the form
\be
p - 1 = -\frac{\rho q_u}{q_0} \, \sin(pq_0L_0) + O(q^b), \label{eq:pa}
\ee
where $b=1/2$ if $\sin(pq_0L_0)=0$ and $b=1$ otherwise.
Also, using $\sn(x,\emod) = \tanh x + O(q)$, we obtain that for large $L-L_0$
\be
\varphi_0(z) = -2\atan\e^{-q_u(z-z_0)}+ O(q). \label{eq:phias}
\ee
Thus, as we said before, $\varphi_0$ has the form of a conventional domain wall centered at $z_0$, with some correction exponentially small with $L$.

Finally, combining equations (\ref{eq:phias}) and (\ref{eq:contphi}) we get 
\be
\tanh\big(q_u(L_0-z_0)\big) = \cos(pq_0L_0) + O(q). \label{eq:contphia}
\ee
This gives an explicit solution for $z_0$ if we neglect the $O(q)$ term.

\begin{figure}[t!]
\includegraphics[width=0.4\textwidth]{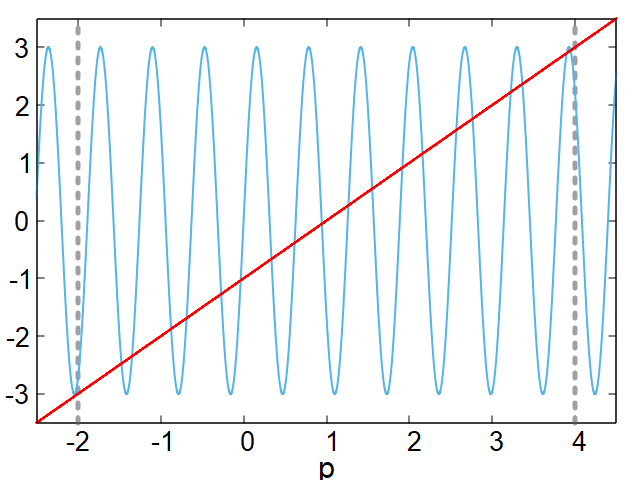}
\includegraphics[width=0.4\textwidth]{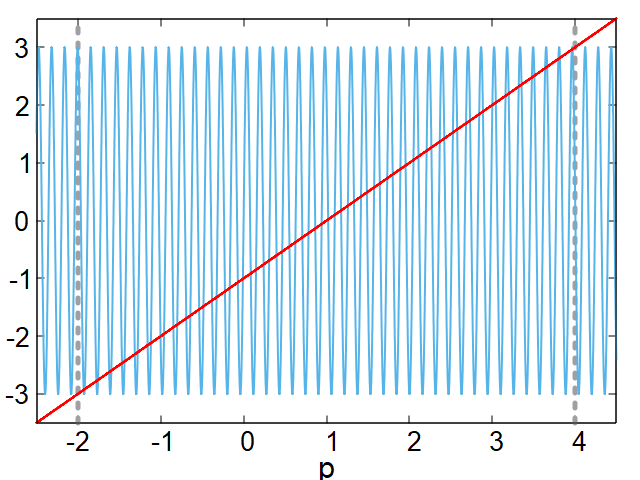}
\caption{Graphical illustration of the solutions of equation (\ref{eq:pa}). The red line is the left-hand side of the equation, and the violet line if the right-hand side. The upper panel corresponds to $q_0L_0=10$ and the lower panel to $q_0L_0=40$. In both cases $\rho q_u/q_0=3$. The vertical dashed lines signal the bounds (\ref{eq:bounds}). \label{fig:p}}
\end{figure}

It is clear that equation (\ref{eq:pa}) has many solutions if we neglect the $O(q)$ term, and it is also clear that this term, exponentially small with $L-L_0$, cannot change this behavior. Moreover,
it is also clear that the number of solutions increases proportionally to $q_0L_0$, and thus the values of $p$ that solve equation (\ref{eq:pa}), or equation (\ref{eq:p}), become dense in the interval (\ref{eq:bounds}) in the limit $L_0\to\infty$. Thus, there is a continuum of helical states in the limit $q_0L_0\to\infty$, as claimed in reference \onlinecite{Laliena2023}. Figure \ref{fig:p} illustrates these statements.

For large $q_0L_0$ the energy density of these helical states, measured from the energy density of the uniform state, coincides with the energy density of the chiral magnet part, since the energy of the uniaxial ferromagnet part remains finite as $q_0L_0$, or $q_uL$, or both, tend to infinity (it is at most the energy of a domain wall). Hence, for $q_0L_0\to\infty$, the energy density of the helical states studied here is given by equation (8) of reference \onlinecite{Laliena2023}, and is shown graphically by figure 1 of this reference. The energy density depends on $p$ and has its minimum at $p=1$. Thus, all the $p$ states with $p\neq 1$, are at most metastable.
It remains to see which, if any, of the $p$ states are actually metastable. This problem is addressed in the next section.

Before closing this section we notice an interesting fact about the solutions of equation (\ref{eq:p}). 
The derivative with respect to $p$ of the right-hand side of equation (\ref{eq:p}) is 
\be
-\rho q_u L_0 \frac{\sin(pq_0L_0) \cos(pq_0L_0)}{\sqrt{\emod^2-\cos^2(pq_0L_0)}}.
\ee
Since the sign of $\sin(pq_0L_0)$ is fixed by $\sigma_p$, see equation (\ref{eq:cont_sin}), the sign of the above expression is determined by $\cos(pq_0L_0)$. It is clear from Fig. \ref{fig:p} that the solutions of equation (\ref{eq:p}) correspond alternatively to points in which the right hand side of equation (\ref{eq:p}) increases and decreases. These means that $\cos(pq_0L_0)$ is positive for one half of the $p$'s and negative for the other one half. Then, equation (\ref{eq:contphia}) implies that one half of the $p$ values correspond to $z_0<L_0$ and the other one half to $z_0>L_0$. In the case $z_0<L_0$ the center of the wall is outside the physical region occupied by the uniaxial ferromagnet, while if $z_0>L_0$ it is within the uniaxial ferromagnet. To lighten the writing, let us call the former case a \textit{virtual} domain wall and the latter a \textit{real} domain wall. Hence, half of the $p$ correspond to virtual domain walls and the other half to real domain walls. We will see in next section that, if $L-L_0$ is large, all $p$ states with real domain wall are unstable and that the $p$ states with a virtual domain wall are stable if $\pmin < p < \pmax$, with
\be
\begin{split}
\pmin = \max\left\{1-\sqrt{\hc},\,1-\frac{\rho q_u}{q_0}\right\}, \\[6pt]
\pmax = \min\left\{1+\sqrt{\hc},\,1+\frac{\rho q_u}{q_0}\right\},
\end{split}
\label{eq:whole_bounds}
\ee 
where $\hc=1-\kappa>1$ is the dimensionless critical field of the chiral magnet. 

The case $p=1$ is somehow especial. In this case equations (\ref{eq:contphi}) and (\ref{eq:pphip}) give $\cos(\varphi_0(L_0))=\pm\emod$. Then, from equation (\ref{eq:cosphi0}), we get $\sn(x_0,\emod)=\pm1$, where $x_0=q_u(L_0-z_0)$. The solution for the plus case is $x_0=K(\emod)$, which, on account of (\ref{eq:K}), is only possible if $L_0=L$, that is, if the size of the uniaxial ferromagnet vanish. So, this possibility is realized only if the system consists of a monoaxial chiral magnet in contact with a nonmagnetic material. For the minus case we have $x_0=-K$, and this gives $z_0=(L_0+L)/2$. This means the center of the wall is in the middle of the uniaxial ferromagnet, hence it is a real wall, and therefore unstable.
In spite of this discussion, $p=1$ can also be realized if the chiral magnet is in contact with a uniaxial ferromagnet, in the sense that there will be metastable states with $p$ arbitrarily close to one if $q_0L_0$ is sufficiently large.

\section{Stability of the helical states \label{sec:stab}}

Let $\hn_p$ be the magnetization of the $p$ state, given by equation (\ref{eq:hn}), with $\varphi$ given by (\ref{eq:phi}). The $p$ state will be stable if it is a local minimum of the energy. 
We ignore the magnetostatic energy, which can only contribute to the stability of the $p$ state.  This is due to the fact that the magnetostatic field created by the $p$ state vanishes, since its sources vanish: $\nabla\cdot\hn_p=0$ and $\hz\cdot\hn_p=0$. Then, the magnetostatic energy of the $p$ state is zero and, since it cannot be negative, any perturbation can only increase it, what means that the magnetostatic interaction cannot contribute to destabilize the $p$ state. 

A necessary condition for the $p$ state to be a local minimum of the energy is the positivity of the second variation of the energy at $\hn_p$. 
To obtain the second variation,
we consider a perturbation of $\hn_p$, which if it is small enough can be written in terms of two fields $\xi_1$ and $\xi_2$ as
\be
\hn = \sqrt{1-\xi^2}\,\hn_p + \xi_1\,\he_1 + \xi_2\,\he_2.
\ee
where we introduce the unit vector fields
\be
\he_1 = -\hz, \quad \he_2 = -\sin\varphi(z)\,\hx + \cos\varphi(z)\,\hy,
\ee
so that $\{\he_1,\he_2,\hn_p\}$ is a right-handed orthonormal triad. To have perturbations of finite energy we restric the fields $\xi_i$, for $i=1,2$, to square integrable functions. The continuity of $\hn$ and the matching conditions at $z=\pm L_0$ provide further conditions for $\xi_i$ and $\partial_z\xi_i$, to be discussed below.

The second variation of the energy, $\delta^{(2)}E$, at $\hn_p$ can be obtained by inserting the above perturbation into the energy functional and expanding in powers of $\xi_1$ and $\xi_2$ to the second order. A straightforward computation gives
\be
\delta^{(2)}E = 2A \int d^3r \, \Big(\xi_1 K_{11}\xi_1 + \xi_2 K_{22}\xi_2\Big),
\ee
where $K_{11}$ and $K_{22}$ are linear differential operators defined by their action on functions $\xi$ as 
\begin{gather}
\hspace{-0.2cm}
K_{11}\,\xi = K_{22}\, \xi + q_0^2\big[\hc-(p-1)^2\big] \chi_c\,\xi, \label{eq:K11}
 \\[6pt]
\hspace{-0.2cm}
K_{22}\,\xi = -a\nabla_T^2\xi - \partial_z(a\partial_z \xi) + \rho \big(q_u^2-2\varphi^{\prime\,2}\big)\chi_u \,\xi, 
\label{eq:K22}
\end{gather}
where, we recall, $\hc=1-\kappa>1$ and $a(z) = \chi_c(z) + \rho\chi_u(z)$ is  the function defined at the end of section \ref{sec:model}.

It is convenient to perform the Fourier transform in the coordinates $x$ and $y$. To avoid symbol proliferation we use the same notation for functions and operators in the real and transformed space.
After the Fourier transformation we have
\be
K_{22}\,\xi =  -\big(a \xi^\prime\big)^\prime + ak_T^2\xi+ \rho \big(q_u^2-2\varphi^{\prime\,2}\big)\chi_u\,\xi,
\ee
and $K_{11}$ is still given by equation (\ref{eq:K11}).
Now $\xi$ is a function of the Fourier wave vector $\vec{k}_T=k_x\hx+k_y\hy$, and of $z$. 

The continuity of $\hn$ and the matching conditions (\ref{eq:matching}) imply that $\xi$ and $a\xi^\prime$ have to be continuous (here $\xi$ represents either $\xi_1$ or $\xi_2$). In particular, the matching condition implies 
\be
 \lim_{z\to L_0^-}\xi^\prime(z) = \lim_{z\to L_0^+}\rho\,\xi^\prime(z). \label{eq:matching_xi}
\ee
An analogous relation holds for $z\to -L_0^{\pm}$. Finally, the boundary conditions for $\hn$ give
\be
\xi^\prime (-L) = 0, \quad \xi^\prime (L) = 0. \label{eq:bc_xi}
\ee

After the Fourier transformation, $K_{11}$ and $K_{22}$ are particular cases of a general kind of differential operators thoroughly studied in reference \onlinecite{Weidmann1987}. Their actions are well defined on continuous functions $\xi$ defined in $[-L,L]$, which are piecewise continuously differentiable and such that $a\xi^\prime$ is also continuous and piecewise continuously differentiable, and satisfy the boundary conditions (\ref{eq:bc_xi}) \footnote{The maximal domain of these operators, where they are selfadjoint, is the set of absolutely continuous functions $u$ such that $au^\prime$ is also absolutely continuous \cite{Weidmann1987}.}. The operators are selfadjoint in the appropiate extended domain \cite{Weidmann1987}.

A necessary condition for the stability of the $p$ state is that the operators $K_{11}$ and $K_{22}$ are positive definite, so that the second variation of the energy is positive definite. This means that their spectrum, which is purely discrete, must lie in the positive real axis. The eigenvalues of $K_{ii}$ are given by the values of $\lambda$ for which the differential equation
\be
K_{ii} \xi = \lambda \xi \label{eq:spectral}
\ee
has solutions which satisfy the matching conditions (\ref{eq:matching_xi}) and the boundary conditions (\ref{eq:bc_xi}). 

\subsection{Bounds on the spectrum}

The spectrum of operators like $K_{11}$ and $K_{22}$, in the definition of which enter only functions bounded from below, is bounded from below. Indeed, multiplying equation (\ref{eq:spectral}) by $\xi$, integrating from $[-L,L]$, and then using integration by parts and the boundary conditions (\ref{eq:bc_xi}), the following bound for the spectrum of $K_{22}$ is obtained:
\be
\lambda \geq \rho (k_T^2-q_u^2). \label{eq:bl2}
\ee
Similarly, for the spectrum of $K_{11}$ we get the bound
\be
\lambda \geq \min\Big\{k_T^2+ \big[\hc-(p-1)^2\big]q_0^2,\;\;\rho (k_T^2-q_u^2)\Big\}. \label{eq:bl1}
\ee

\subsection{Eigenvalue equations}

To study the spectrum it is convenient to introduce the quantities
\begin{gather}
\beta = 1 + \frac{k_T^2-\lambda/\rho}{q_u^2}, \label{eq:beta} \\[4pt]
\gamma_1 = \gamma_2 + \hc-(p-1)^2,  \\[4pt]
\gamma_2 = \rho \left(\beta - 1 - \frac{k_T^2}{q_u^2}\right) \frac{q_u^2}{q_0^2} + \frac{k_T^2}{q_0^2}.
\end{gather}

Since the operators $K_{11}$ and $K_{22}$ commute with the parity operator, their eigenfunctions can be chosen as even or odd functions.
From the form of these operators, we see the that the eigenfunctions, $u$, of $K_{ii}$, with $i=1,2$, can be written, for $z\geq 0$, as
\be
u(z) = c_1 v(q_0z,\gamma_i) \chi_c(z) + c_ 2 w\big(x,\beta\big)\chi_u(z), \label{eq:evec}
\ee
where $x=q_u(z-z_0)$, $c_1$ and $c_2$ are constants to be determined,  $v(x,\gamma)$ is a particular, even or odd, solution of
\be
v^{\prime\prime} - \gamma v = 0, \label{eq:ui}
\ee
and $w(x,\beta)$ is a particular solution of
\be
w^{\prime\prime}-2\varphi_0^{\prime\,2}  w - \beta w = 0, 
\label{eq:uo}
\ee
which satisfies the condition $w^\prime\big(K,\beta\big)=0$. Equations (\ref{eq:ui}) and (\ref{eq:uo}) are simply the restriction of equation (\ref{eq:spectral}) to $|z|<L_0$ and $L_0<z<L$, respectively. In equation (\ref{eq:uo}) we use the coordinate $x=q_u(z-z_0)$, and thus $z=L$ corresponds to $x=K$, due to equation (\ref{eq:K}). The form of the eigenfunction for $z<0$ can be obtained from the parity symmetry.
It should be clear that, in equation (\ref{eq:uo}), $\varphi_0^{\prime\,2}$, which is given by equation (\ref{eq:phip2}) of appendix \ref{app:sg}, is evaluated at $z=z_0+x/q_u$. 

We are interested only in studying the existence of non positive eigenvalues, $\lambda\leq 0$. For this case equations (\ref{eq:bl2}) and (\ref{eq:bl1}) give the bounds
\be
\beta_\text{min} \leq \beta \leq \beta_\text{max},
\ee
where $\beta_\text{min}=1+k_T^2/q_u^2$ for both $K_{11}$ and $K_{22}$, while for $K_{11}$ we have
\be
\beta_\text{max}=\max\left\{2, 1+\frac{\rho-1}{\rho}\,\frac{k_T^2}{q_u^2} +\frac{q_0^2}{\rho q_u^2}\left(\hc-(p-1)^2\right)\right\},
\ee
and for $K_{22}$ we have $\beta_\text{max}=2$.

The functions $v$ and $w$ entering equation (\ref{eq:evec}) have to fullfill the matching conditions at $z=L_0$ (then, the parity symmetry guaranties that they are fullfilled also at $z=-L_0$). This conditions can hold non trivially (that is, with $u\neq 0$) only for specific values of $\beta$, which give the eigenvalues of the corresponding operator. 

We observe that we have four matching conditions, corresponding to the even and odd eigenfunctions of $K_{11}$ and $K_{22}$. We identify each condition by a pair $(i,s)$, where $i=1,2$ and $s=e,o$ label the operator and the eigenfunction parity, respectively. Each matching condition sets a system of two homogeneous linear equations where the unkowns are the constants $c_1$ and $c_2$ entering equation (\ref{eq:evec}). To have non trivial solutions a condition
\be
F_i^{(s)}(\beta) = 0 \label{eq:evals}
\ee
must hold. The four functions $F_i^{(s)}(\beta)$ are given by
\be
F_i^{(s)}(\beta) = v_s^\prime(q_0L_0,\gamma_i) w(x_0,\beta) - \frac{\rho q_u}{q_0} 
v_s(q_0L_0,\gamma_i) w^\prime(x_0,\beta) \label{eq:eval}
\ee
where $x_0=q_u(L_0-z_0)$, and $v_s(x,\gamma)$ are even and odd solutions of equation (\ref{eq:ui}), which can be chosen as
\bea
&&\hspace*{-0.5cm}
v_e(x,\gamma) = \cosh(\sqrt{\gamma}x), \; v_o(x,\gamma) = \sinh(\sqrt{\gamma}x), \;   \gamma>0,
\nonumber \\[4pt]
&&\hspace*{-0.5cm}v_e(x,\gamma) = 1, \quad v_o(x,\gamma) = x, \quad   \gamma=0, \\[4pt]
&&\hspace*{-0.5cm}v_e(x,\gamma) = \cos(\sqrt{-\gamma}x), \; v_o(x) = \sin(\sqrt{-\gamma}x), \;   \gamma<0.
\nonumber
\eea

For given $p$, equations (\ref{eq:evals}) determine the eigenvalues of $K_{11}$ and $K_{22}$. The $p$ state will be stable if none of these equations has solutions with $\beta$ between $\beta_\text{min}$ and $\beta_\text{max}$.

\subsection{Solution of equation (\ref{eq:uo})}

It remains to find the solutions of equation (\ref{eq:uo}), which is studied in appendix \ref{app:lame}. Here, we summarize the results. The solutions which satisfies the boundary condition $w^\prime(K,\beta)=0$ can be written as
\be
w(x,\beta) = w_+(x,\alpha) + d\, w_-(x,\alpha), \label{eq:sol_lame}
\ee
where $w_+$ and $w_-$ are two linearly independent solutions of equation (\ref{eq:uo}) which can be expressed in terms of the Jacobi theta functions $\theta_1$ and $\theta_2$ as
\be
w_\pm(x,\alpha) = \pm \frac{\phi_1^\prime(0,q)}{\phi_1(\alpha,q)}\,
\frac{\phi_2(x\pm\alpha,q)}{\phi_2(x,q)} \exp
\left(\mp \frac{\phi_1^\prime(\alpha,q)}{\phi_1(\alpha,q)}\,x\right) \label{eq:wpm}
\ee
with
\be
\phi_i(x,q) = \theta_i\left(\iu\frac{\pi x}{2\bar{K}},q\right), \quad i=1,2,3,4. \label{eq:thetaphi}
\ee
The nome $q$ is defined just above equation (\ref{eq:KKb}).
The parameter $\alpha>0$, which has nothing to do with the Gilbert damping parameter entering the LLG equation, is related to $\beta$ through the equation
\be
\sn^2(\alpha,\emod) = \frac{1}{\beta + 1-\emod^2}, \label{eq:alpha_sn}
\ee
and the constant $d$ is determined from the boundary condition, $w^\prime(K,\beta)=0$, and is given by
\be
d = -\exp\left(\frac{\pi\alpha}{\bar{K}} - 2\frac{\phi_1^\prime(\alpha,q)}{\phi_1(\alpha,q)}K\right). \label{eq:d}
\ee

Now we can introduce $w(x,\beta)$ in the eigenvalue equations (\ref{eq:eval}) and analyze them numerically. However, it is useful to analyze first the limit of large $q_u(L-L_0)$, which leads to important simplifications.

\subsection{Analysis for large $q_u(L-L_0)$ \label{sec:large_L}}

The large $q_u(L-L_0)$ regime corresponds to $q\to0$. This limit is studied in appendix \ref{app:qexp}, where the formulas used in this section are derived. There, it is observed that we have to distinguish the case $\beta>1$ from the case $\beta=1$, which is especial.

If $\beta$ is not too close to one the solution of equation (\ref{eq:alpha_sn}) is
\be
\alpha = \sqrt{\beta} + O(q),
\ee
and, taking into account that $K=-\log\sqrt{q}+O(q\log q)$, we have
\be
d = -\exp\Big(2\alpha+\sqrt{\beta}\log q+O(q\log q)\Big). \label{eq:das}
\ee
Hence $d$ is negligible for large $L$, which corresponds to small $q$. Then $w(x,\beta)$ can be approximated by $w_+(x,\alpha)$, which in its turn can be expanded in powers of $q$. 
For fixed $x$ we obtain the simple expression
\be
w(x,\beta) = \left(\sqrt{\beta}+\tanh x\right)\,\e^{-\sqrt{\beta}x} + O(q\log q). \label{eq:wgt1}
\ee
Neglecting the $O(q\log q)$ terms, this is the solution we would have obtained had we considered an infinite system with $u^\prime(x,\beta)\to 0$ for $x\to\infty$ as a boundary condition.

We now analyze $F_2^{(e)}(\beta)$ for $k_T=0$ and large $\beta$. Using equation (\ref{eq:wgt1}), ignoring the $O(q\log q)$ corrections, we obtain that for $\beta\to\infty$
\be
F_2^{(e)}(\beta) \sim \left(1+\frac{1}{\sqrt{\rho}}\right)\frac{\rho q_u}{q_0}
\exp\Big(\sqrt{\beta}\big(\sqrt{\rho}q_uL_0-x0\big)\Big).
\ee
Thus, $F_2^{(e)}(\beta)>0$ for large $\beta$.

For $\beta=1$ the solution of equation (\ref{eq:alpha_sn}) has the form $\alpha=K-\bar{\alpha}$, where $\bar{\alpha}$ is of order one for $q\to0$ (see appendix \ref{app:qexp}). The leading term in $q$ of $w(x,\beta=1)$ is given by equation (\ref{eq:wbeta1}). From it, it is straightforward to get 
\be
F^{(e)}_2(1) = \big(6+2\sqrt{6}\big) \frac{\rho q_u}{q_0}\frac{\tanh x_0}{\cosh x_0} + O(q\log q).
\ee
Thus, $F^{(e)}_2(1)<0$ if $x_0<0$. Therefore, the corresponding states are unstable because $F_2^{(e)}(\beta)$ has a zero for $\beta>1$. Since $x_0<0$ means $z_0>L_0$, we see that all $p$ states which have a real domain wall are unstable, as claimed in section \ref{sec:p}. These unstable states correspond to half of the solutions of equation (\ref{eq:p}).

\subsection{Analysis of states with large $p$ \label{sec:large_p}}

Consider again $k_T=0$. If $|p-1|>\sqrt{\hc}$ we have that $\gamma_1<0$ in a neighborhood of $\beta=1$. Then, for $\beta$ sufficiently close to one we have
\begin{gather}
F_1^{(e)}(\beta) = -\sqrt{-\gamma_1}\sin\big(\sqrt{-\gamma_1}q_0L_0\big) \,w(x_0,\beta) 
\nonumber \\[4pt]
-\frac{\rho q_u}{q_0} \sin\big(\sqrt{-\gamma_1}q_0L_0\big) \,w^\prime(x_0,\beta).
\end{gather}
By continuity, small changes of $\beta$ produce small changes on $w(x_0,\beta)$, $w^\prime(x_0,\beta)$, and $\gamma_1$. But if $q_0L_0$ is large, the trigonometric functions entering the above equation suffer big oscillations, so that that $F_1^{(e)}(\beta)$ changes sign in a neighborhood of $\beta=1$. Hence, states with $|p-1|>\sqrt{\hc}$ are unstable. This relation, toghether with equation (\ref{eq:bounds}), provides the bounds (\ref{eq:whole_bounds}).

\section{Discusion of some results \label{sec:results}}

Let us discuss the results for two systems: one in which the uniaxial ferromagnets attached to the chiral magnet are thick and another one in which they are very thin. In both cases we consider $\hc=6$, $\rho=3$ and $q_u=q_0$. The possible $p$ states are obtained by solving numerically the coupled equations (\ref{eq:K}) and (\ref{eq:p}), and their stability by evaluating the functions $F_i^{(s)}(\beta)$ numerically.

\subsection{Thick slabs of uniaxial ferromagnets}

We take $q_u(L-L_0)=40$, which can be considered in the large $q_u(L-L_0)$ regime, and thus we confirm by numerical means the results of section \ref{sec:stab}. 
We get the following results:
\begin{enumerate}
\item States which have a real domain wall are unstable, whatever the value of $p$, in agreement with the analysis of section \ref{sec:large_L}.
\item States with $|p-1|>\sqrt{\hc}$ are always unstable, in agreement with the argument of section \ref{sec:large_p}.
\item States with $|p-1|<\sqrt{\hc}$ and with a virtual domain wall are metastable. 
\item The number of metastable $p$ states, $N_p$, grows linearly with $q_0L_0$ (see table \ref{tab:xL40}), and the values of such $p$ are homogeneously distributed in the interval $1-\sqrt{\hc}\leq p \leq 1+\sqrt{\hc}$. Hence, the bounds (\ref{eq:whole_bounds}) are saturated.
\end{enumerate}
All this conclusions are in agreement with the analysis of the large $q_u(L-L_0)$ regime presented in section \ref{sec:large_L}, and imply that in the limit $q_0L_0\to\infty$ the results of reference \onlinecite{Laliena2023} are recoverd.

\begin{figure}[t!]
\includegraphics[width=0.23\textwidth]{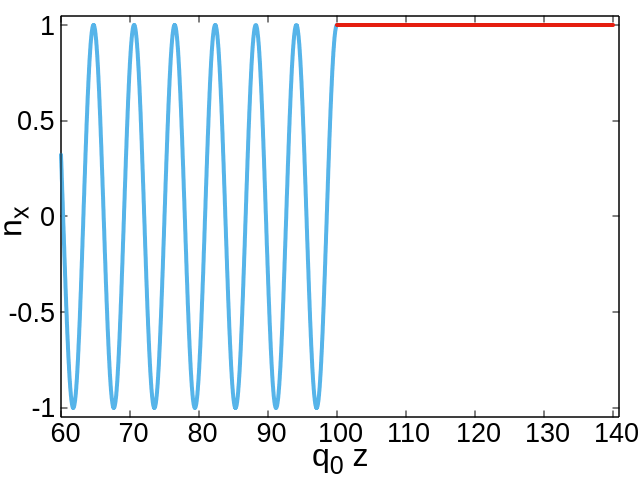}
\includegraphics[width=0.23\textwidth]{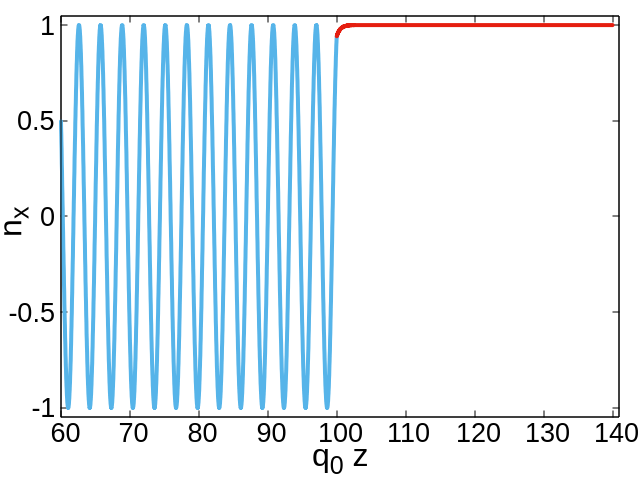}
\includegraphics[width=0.23\textwidth]{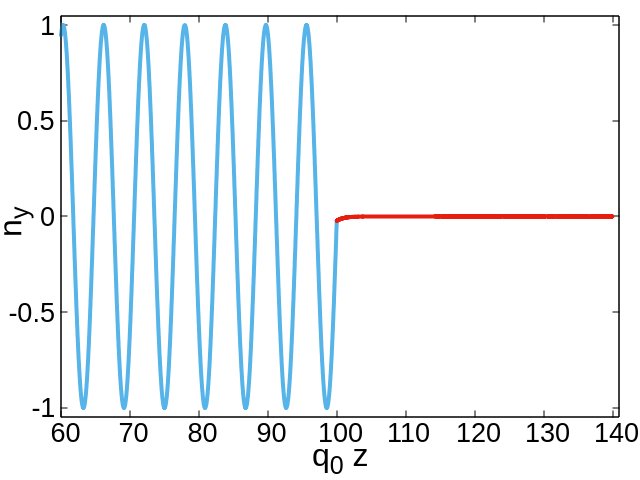}
\includegraphics[width=0.23\textwidth]{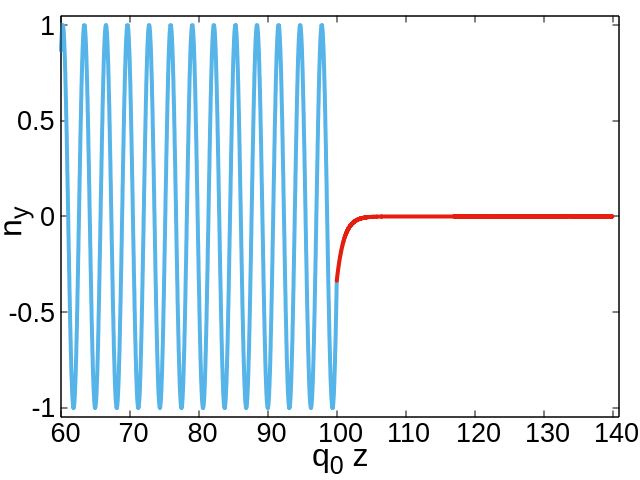}
\includegraphics[width=0.23\textwidth]{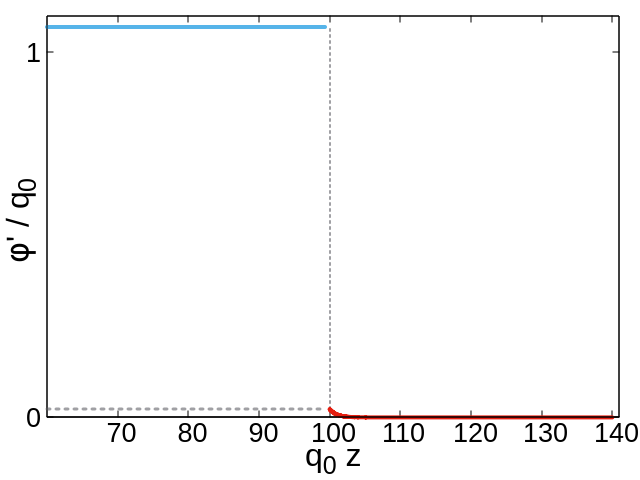}
\includegraphics[width=0.23\textwidth]{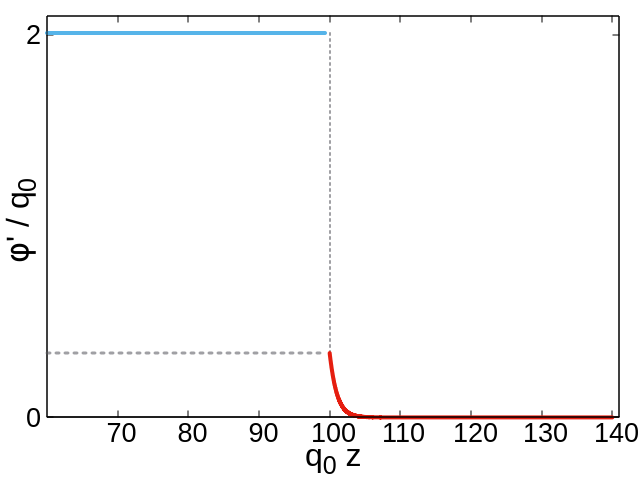}
\caption{Magnetization of $p$ states in the composite magnet system for the case of thick uniaxial ferromagnet slabs, with $q_u(L-L_0)=40$. The size of the chiral magnet is $q_0L_0=100$. The system parameters are given at the beginning of section \ref{sec:results}.
From top to bottom the figures show $n_x$, $n_y$, and $\varphi^\prime/q_0$ versus $q_0z$.
The left panels correspond to $p=1.068$ and the right panels to $p=2.007$. 
 \label{fig:magnet_xL40}}
\end{figure}

Figure \ref{fig:magnet_xL40} shows the magnetization in two representative cases, one for $p=1.068$ (the closest value to $p=1$) and another one for $p=2.007$ (the closest value to $p=2$). Observe that the magnetization in the uniaxial ferromagnet region has the form a virtual domain wall, again in agreement with what we expected after the analysis of the large $q_u(L-L_0)$ regime.
The bottom panels show the derivative $\varphi^\prime(z)/q_0$. The discontinuity at $z=100$ is due to the matching condition (\ref{eq:pphip}).

\begin{table}[t!]
\begingroup
\setlength{\tabcolsep}{4pt} 
\renewcommand{\arraystretch}{1.5} 
\begin{tabular}{c|ccccccc}
\hline\hline
$q_0L_0$ & 10 & 20 & 40 & 60 & 80 & 100 & 120 \\ [1pt]
\hline
$N_p$ & 8 & 16 & 32 & 47 & 63 & 78 & 93 \\ [1pt]
\hline
$N_p/q_0L_0$ & 0.8 & 0.8 & 0.8 & 0.78 & 0.79 & 0.78 & 0.78 \\
\hline\hline
\end{tabular}
\endgroup
\caption{Number of metastable states, $N_p$, \textit{versus} $q_0L_0$ for the case of thick slabs of uniaxial ferromagnet: $q_u(L-L_0)=40$. The system parameters are given at the beginning of section \ref{sec:results}.
\label{tab:xL40}}
\end{table}

\subsection{Thin slabs of uniaxial ferromagnets}

To analyze the case in which the uniaxial ferromagnets are very thin slabs we take $q_u(L-L_0)=1$, which means that the thickness of the uniaxial ferromagnets is equal to the domain wall width of a very thick magnet. Table \ref{tab:xL1} shows that the number of metastable $p$ states is proportional to $q_0L_0$. The values of $p$ corresponding to metastable states are homogeneously distributed in the interval $-0.43<p<2.44$,  what means that there is a continuum of metastable $p$ states in the limit $q_0L_0\to\infty$. Notice that the interval of metastable states is within the limits (\ref{eq:whole_bounds}), but does not saturate them. However, these limits are rapidly saturated by increasing the thickness of the uniaxial ferromagnets, $q_u(L-L_0)$. Indeed, with a thickness equal to twice the domain wall width, $q_u(L-L_0)=2$, the bounds are already saturated.

\begin{table}[t!]
\begingroup
\setlength{\tabcolsep}{4pt} 
\renewcommand{\arraystretch}{1.5} 
\begin{tabular}{c|ccccccc}
\hline\hline
$q_0L_0$ & 10 & 20 & 40 & 60 & 80 & 100 & 120 \\ [1pt]
\hline
$N_p$ & 10 & 19 & 38 & 56 & 76 & 92 & 112  \\ [1pt]
\hline
$N_p/q_0L_0$ & 1 & 0.95 & 0.95 & 0.93 & 0.95  & 0.92 & 0.93 \\
\hline\hline
\end{tabular}
\endgroup
\caption{Number of metastable states, $N_p$, \textit{versus} $q_0L_0$ for the case of thin slabs of uniaxial ferromagnets: $q_u(L-L_0)=1$. The system parameters are given at the beginning of section \ref{sec:results}.
\label{tab:xL1}}
\end{table}

\begin{figure}[t!]
\includegraphics[width=0.23\textwidth]{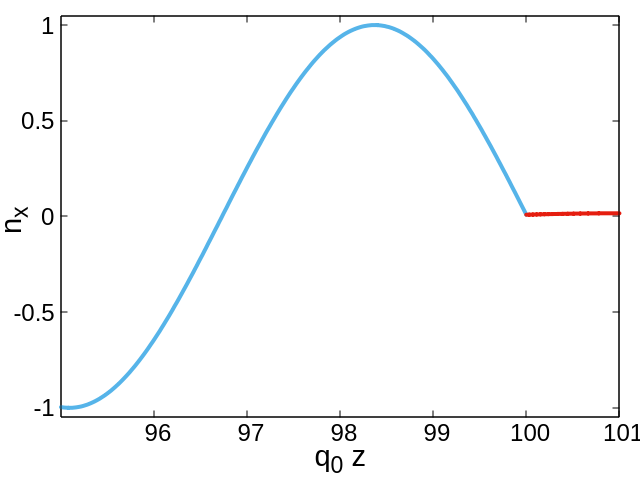}
\includegraphics[width=0.23\textwidth]{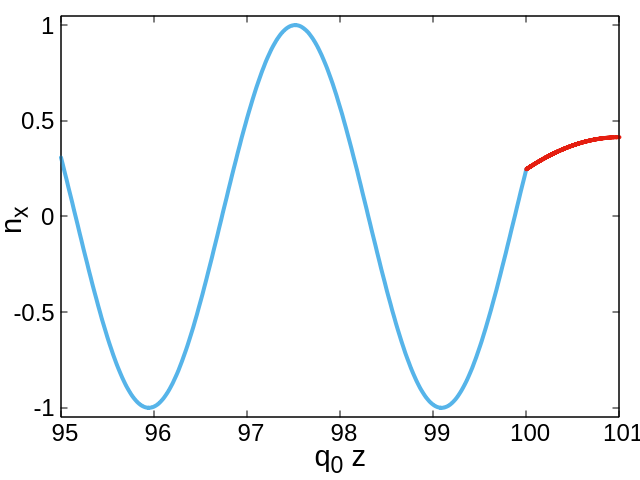}
\includegraphics[width=0.23\textwidth]{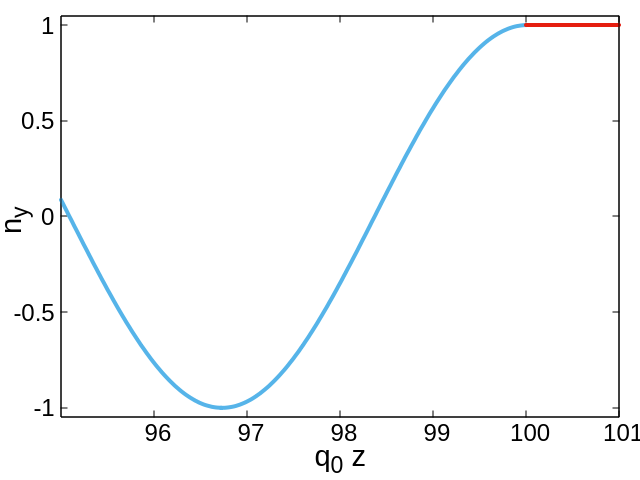}
\includegraphics[width=0.23\textwidth]{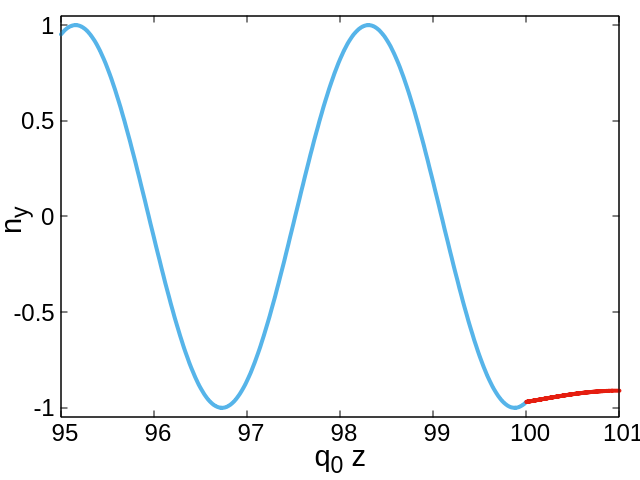}
\includegraphics[width=0.23\textwidth]{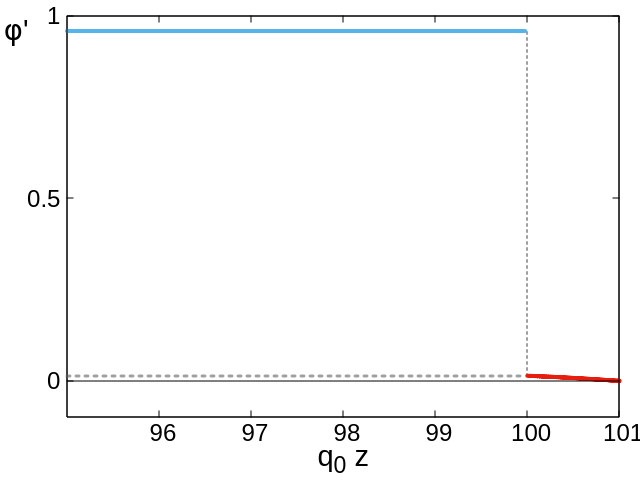}
\includegraphics[width=0.23\textwidth]{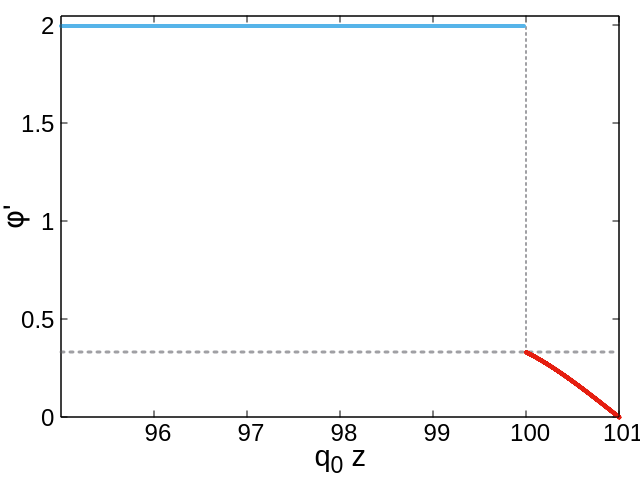}
\caption{Magnetization of $p$ states in the composite magnet system for the case of thin uniaxial ferromagnet slabs, with $q_u(L-L_0)=1$. 
The size of the chiral magnet is $q_0L_0=100$. The system parameters are given at the beginning of section \ref{sec:results}.
From top to bottom the figures show $n_x$, $n_y$, and $\varphi^\prime/q_0$ versus $q_0z$.
The left panels correspond to $p=0.958$ and the right panels to $p=1.997$. 
 \label{fig:magnet_xL1}}
\end{figure}

For illustration, figure \ref{fig:magnet_xL1} shows the magnetization components and the derivative $\varphi^\prime(z)$ for the cases $p=0.958$ (the closest value to $p=1$) and $p=1.997$ (the closest value to $p=2$). We see that the magnetization in the thin uniaxial ferromagnet does not look like a section of a virtual domain wall (for instance, $n_y$ should tend to zero in the case of a domain wall, but it does not in the prersent case). The bottom panels show the derivative $\varphi^\prime(z)/q_0$. The discontinuity at $z=100$ is due to the matching condition (\ref{eq:pphip}).

\section{Conclusions \label{sec:conc}}

In reference \onlinecite{Laliena2023} we showed that in an infinite monoaxial chiral magnet there exists a continuum of metastable helical states differing by the helix wave vector, $pq_0$. It was pointed out that in a real finite magnet only the state with $p=1$ is compatible with the surface chiral twist induced by the natural boundary conditions \cite{Garst_private}. This means that states with $p\neq 1$ are ruled out by the boundary conditions, and, apparently, the results of reference \onlinecite{Laliena2023} only hold in the non physical cases of an infinite magnet or of a magnet with periodic boundary conditions. 

However, the boundary conditions can be taylored by attaching some other magnet to the faces of the monoaxial chiral magnet which are perpendicular to the chiral axis. These magnets may absorb the chiral twist and thus states with different $p$ may satisfy the boundary conditions. We proof in this work that this is indeed the case by considering a composite magnet system formed by a monoaxial chiral magnet attached to two similar uniaxial ferromagnets, as in figure 1. In this work we deal only with the case of zero applied field and zero applied current, since this problem can be solved exactly. The cases of non zero applied field and/or nonzero applied current are very interesting, but have to be addressed numerically.

For zero applied field, we show here that if the ferromagnets are thick enough (thickness a few tens times larger than the width of their characteristic domain wall) this composite system has metastable magnetic states which are helical within the chiral magnet and look like a virtual domain wall within the ferromagnets (by virtual we mean that the center of the wall is outside the physical region occupied by the uniaxial ferromagnets). Those metastable states differ by the wave number of the helix within the chiral magnet and its number increases by increasing the size of the chiral magnet, $L_0$. The results of reference \onlinecite{Laliena2023} are thus fully recovered in the limit $L_0\to\infty$. 

We also obtain results similar to those of reference \onlinecite{Laliena2023} (for zero applied field) in the limit $L_0\to\infty$ if the ferromagnets attached to the chiral magnet are thin (thickness approximately equal to the domain wall width). In this case, however, there is a difference, since the range of $p$ for which the helical states are metastable is shorter than that predicted in  \onlinecite{Laliena2023}. 
 
At first sight, the existence of so many non degenerate metastable states in chiral magnets is somehow disconcerting, since we usually believe that the boundary conditions select one of the many solutions of the differential equations that govern the dynamics of the system in the static case. In this case, however, if the temperature is low enough and if the energy barriers between the $p$ states are high enough, it is the initial condition what determines the magnetic state in the long term.
Actually, this situation resembles the physics of ferromagnets, in which the existence of many metastable states characterized by different spatial distributions of domains is at the origin of hysteresis. 

To conclude, let us stress that the possible uses of the $p$ states depend strongly on their life times, which in turn depend on the energy barriers which separate them. If the barriers are high enough, the $p$ states could be experimentally detected at low enough temperature in a system similar to that studied in this work.

\begin{acknowledgments}
We thank M. Garst and J. Masell for pointing out the issues with the boundary conditions and for useful discussions.
Grants No. PID2022-138492NB-I00-XM4, funded by MCIN/AEI/10.13039/501100011033, and E11\_23R/M4, funded by Diputaci\'on General de Arag\'on, supported this work.  
This work was also supported by the Grant No. PICT 2017-0906 from the Agencia Nacional de Promoci\'on Cient\'ifica y Tecnol\'ogica, Argentina.
\end{acknowledgments}
	
\appendix

\section{A solution of the double Sine-Gordon equation \label{app:sg}}

The solution of equation (\ref{eq:sg}) which satisfy the conditions (\ref{eq:condphi0}) can be obtained as follows.
Multiplying equation (\ref{eq:sg}) by $\varphi^\prime$ we get
\be
\frac{d}{dz}\left(\varphi^{\prime\,2}-q_u^2\sin^2\varphi\right) = 0,
\ee
so that the term within brackets has to be a constant. Since we require $\varphi^\prime(L)=0$ the constant has to be 
$-q_u^2\sin^2\varphi_0(L)$ and we obtain
\be
\varphi^\prime = q_u\sqrt{\sin^2\varphi - \sin^2\varphi_0(L)}, \label{eq:phip}
\ee
since we also require $\varphi^\prime>0$. Let us call $\emod=\cos\varphi_0(L)$. Notice that $0<\emod<1$.
Then, solving the above differential equation by separation of variables, we have
\be
\int_{-\frac{\pi}{2}}^{\varphi_0} \frac{d\varphi}{\sqrt{\emod^2-\cos^2\varphi}} = q_u(z-z_0).
\ee
With the change of variable $\emod\, t = \cos\varphi$ in the integral, so that $\sin\varphi=-\sqrt{1-\emod^2t^2}$, we obtain
\be
\int_0^{\frac{1}{\emod}\cos\varphi_0}\!\!\!\! \frac{dt}{\sqrt{(1-t^2)(1-\emod^2 t^2)}} = q_u(z-z_0).
\label{eq:int2}
\ee
The integral of the left-hand side is 
\be
\mathrm{arcsn}\big(\cos\varphi_0/\emod,\emod\big),
\ee
where $\mathrm{arcsn}(x,\emod)$ is the inverse Jacobi elliptic function \cite{Gradshteyn2007}.
Hence, we get
\be
\cos\varphi_0 = \emod\,\sn\big(q_u(z-z_0),\emod\big). \label{eq:cosphi0}
\ee
The parameter $\emod$ is determined by setting $z=L$ in equation (\ref{eq:int2}), in which case $\cos\varphi_0=\eta$ and the upper limit of the integral (\ref{eq:int2}) is one. Thus the integral becomes the complete elliptic integral of the first kind, $K\big(\emod\big)$, so that
\be
K\big(\emod\big) = q_u(L-z_0).
\ee
The above equation determines uniquely $\emod$ if $L$ and $z_0$ are given. Then, since $-\pi<\varphi_0<0$, equation 
(\ref{eq:cosphi0}) determines completely $\varphi_0$, which is given by equation (\ref{eq:phi0}). In particular, since $\sin\varphi_0<0$, we have
\be
\sin\varphi_0 = - \sqrt{1-\cos^2\varphi_0}.  \label{eq:sinphi0}
\ee
We also need an explicit form of $\varphi_0^{\prime\,2}$, which, taking into account (\ref{eq:phip}) and (\ref{eq:cosphi0}), has the form
\be
\varphi_0^{\prime\,2}(z) = q_u^2\eta^2\Big(1-\sn^2\big(q_u(z-z_0),\emod\big)\Big). \label{eq:phip2}
\ee

\section{Solution of equation (\ref{eq:uo}) \label{app:lame}}

In this appendix we outline a way of solving  equation (\ref{eq:uo}) which relies on the Weierstrass elliptic function $\wp$ with fundamental half periods chosen as $\omega_1=\iu \bar{K}$ and $\omega_3=-K$. We use the notation of reference \onlinecite{DLMF} for the fundamental half periods, and $K$ and $\bar{K}$ are defined in equation (\ref{eq:KKb}). This choice of fundamental half periods gives the nome $q=\exp(-\pi K/\bar{K})$, which is convenient if $L$ is large.
The related Weierstrass functions $\xi$ and $\sigma$ and the Jacobi theta functions also appear in the solution. The properties of these functions are thoroughly presented for instance in references \onlinecite{Whittaker1927} or \onlinecite{Akhiezer1990}, and an exhaustive summary can be found in reference \onlinecite{DLMF}. It should be clear that the Weierstrass function $\xi$ of this appendix has nothing to do with the functions $\xi_1$, $\xi_2$, and $\xi$ of section \ref{sec:stab}.

Using equation (\ref{eq:phip}) from appendix \ref{app:sg}, we see that equation (\ref{eq:uo}) has the form
\be
u^{\prime\prime} -2\emod^2\sn^2(x,\emod) - \big(\beta-2\emod^2\big) u = 0, 
\ee
This is one of Lame's equation in Jacobian form \cite{Arscott1964}. Expressing $\sn(x,\emod)$ in terms of $\wp$, the equation is cast to the form
\be
w^{\prime\prime}-2\wp\big(x+\iu\bar{K}\big) w - \left(\beta - \frac{2}{3} + \frac{4}{3}(1-\emod^2)\right)w = 0.
\ee
For given $\beta$, its general solution \cite{Arscott1964} is a linear combination 
\be
w(x,\beta) = d_1 w_+(x,\alpha) + d_2 w_-(x,\alpha),
\ee
where $d_1$ and $d_2$ are arbitrary constants, 
\be
w_\pm(x,\alpha) = \pm \frac{\sigma(x+\iu\bar{K}\pm\alpha)}{\sigma(x+\iu\bar{K})\sigma(\alpha)}
\,\e^{\pm\xi(\alpha)x}. \label{eq:wpm_app}
\ee
and $\alpha$ is the solution of
\be
\wp(\alpha) = \beta-\frac{2}{3} + \frac{4}{3}(1-\emod^2),
\ee
which, using again the relation between $\sn(x,\emod)$ and $\wp(x)$, leads to equation (\ref{eq:alpha_sn}).

If $w^\prime_-(K,\alpha)\neq 0$, the boundary condition $w^\prime(K)=0$ gives 
\be
\frac{d_2}{d_1} = -\frac{w^\prime_+(K,\alpha)}{w^\prime_-(K,\alpha)}. \label{eq:d2d1}
\ee

It is convenient to express $\sigma$ and $\xi$ in terms of theta functions, using equations 23.6.9 and 23.6.13 of reference \onlinecite{DLMF}, since these functions have Fourier series rapidly convergent for small $q$.
We get
\be
\begin{split}
 \frac{\sigma(x+\iu\bar{K}\pm\alpha)}{\sigma(x+\iu\bar{K})\sigma(\alpha)} & = 
 \frac{\phi_1^\prime(0,q)\,\phi_1(x+i\bar{K}\pm\alpha,q)}{\phi_1(x+i\bar{K},q)\,\phi_1(\alpha,q)} \,\times
  \\[2pt]
  & \exp\left(\pm\frac{\xi(\omega_1)}{\omega_1}\alpha(x+\iu\bar{K})\right),
  \label{eq:sigma_app}
\end{split}
\ee
\be
\xi(\alpha) = \frac{\xi(\omega_1)}{\omega_1}\alpha + \frac{\phi_1^\prime(\alpha,q)}{\phi_1(\alpha,q)},
\label{eq:xi_app}
\ee
where the functions $\phi_i(z,q)$ are related to theta functions by equation (\ref{eq:thetaphi}). 
We also use the periodicity of the theta functions (formulas 20.2.6 and 20.2.12 of reference \onlinecite{DLMF}) to obtain
\be
\phi_1\big(z+\iu\bar{K}\big) = -\phi_2(z,q). \label{eq:phi_app}
\ee
Inserting equations (\ref{eq:sigma_app}), (\ref{eq:xi_app}) and (\ref{eq:phi_app}) into equation (\ref{eq:wpm_app}), and removing the factor $\exp\big(\mp\xi(\omega_1)\alpha\big)$, which is a pure phase factor of order one as $q\to0$, what amounts merely to a redefinition of $w_\pm(x,\alpha)$, we obtain equation (\ref{eq:wpm}). 

The derivatives of $w_\pm(x,\alpha)$ can be readily computed from equation (\ref{eq:wpm}), obtaining
\be
\frac{w_\pm^\prime(x,\alpha)}{w_\pm(x,\alpha)} = \pm\frac{\phi_2^\prime(x\pm\alpha,q)}{\phi_2(x\pm\alpha,q)}
\pm\frac{\phi_2^\prime(x,q)}{\phi_2(x,q)} \mp \frac{\phi_1^\prime(\alpha,q)}{\phi_1(\alpha,q)}.
\ee

To compute $d_2/d_1$, equation (\ref{eq:d2d1}), we have to evaluate $w_\pm^\prime\big(K,\alpha\big)$, for which we use the behaviour of theta functions under translation by half periods, given by equation 20.2.13 of reference \onlinecite{DLMF}. Taking into account that $\theta_3^\prime(0,q)=0$ and setting $d_1=1$ and $d_2=d$ we arrive at equation (\ref{eq:K}).

\section{Expansion in $q$ for large $L$ \label{app:qexp}}

\subsection{Case $\beta>1$}

For $q\to0$ and $\beta>1+c$, where $c>0$ is any fixed number, independent of $q$, equation (\ref{eq:alpha_sn}) can be expanded in powers of $q$, by introducing the expansion $\alpha=\alpha_0+\alpha_1 q + \ldots$. Since $\sn(\alpha,\emod)=\tanh(\alpha)+O(q)$, we get for the leading order
\be
\alpha_0 = \atanh\big(1/\sqrt{\beta}\big).
\ee
It is clear that this expansion is not valid for $\beta\to 1$, since in this limit $\alpha_0\to\infty$.

For $q\to0$ we have $K=-\log(\sqrt{q})+O(q\log q)$ and 
\be
\frac{\pi\alpha}{\bar{K}} = 2\alpha + O(q), \quad 
\frac{\phi_1^\prime(\alpha,q)}{\phi_1(\alpha,q)} = \sqrt{\beta}+ O(q). \label{eq:qexp_r1}
\ee
Inserting these equations into equation (\ref{eq:d}) we arrive at equation (\ref{eq:das}), and we see that $d$ vanishes exponentially as $q_uL\to\infty$. Hence, $w(x,\beta)$ can be approximated by $w_+(x,\alpha)$ in this limit. The expansion of this function in powers of $q$ is obtained from
\be
\frac{\phi_1^\prime(0,q)}{\phi_1(\alpha,q)} = 1+O(q), \quad
\frac{\phi_2(x+\alpha,q)}{\phi_2(x,q)} =\sqrt{\beta}-\tanh x +O(q), 
\ee
and from the second of equations (\ref{eq:qexp_r1}). Inserting these equations into equation (\ref{eq:wpm}) for $w_+(x,\alpha)$ we obtain equation (\ref{eq:wgt1}). 

\subsection{Case $\beta=1$}

For $\beta=1$ the expansion in power of $q$ is different. In this case the right hand side of equation (\ref{eq:alpha_sn}) is $1+O(q)$. Taking into account that $\sn(K,\emod)=1$, we see that the solution has the form $\alpha=K-\bar{\alpha}$, where $\bar{\alpha}$ is of order one as $q\to0$. Using the relation
$\sn(K-\bar{\alpha},\emod)=\cd(\bar{\alpha},\emod)$, where $\cd$ is the ratio of the $\cn$ and $\dn$ Jacobi elliptic functions, we get the equation for $\bar{\alpha}$, 
\be
\cd^2(\bar{\alpha},\emod) = 1 - \frac{1-\emod^2}{2-\emod^2}.
\ee
Now we can expand this equation in powers of $q$, inserting the expansion $\bar{\alpha}=\bar{\alpha}_0+\bar{\alpha}_1q+\ldots$. Using
\bd
\cd(\bar{\alpha},q)= 1 -4\big(\cosh(2\bar{\alpha})-1\big)\,q + O(q^2),
\ed
we obtain $\bar{\alpha}_0=\asinh(\sqrt{2})$.

Then we insert $\alpha=K-\bar{\alpha}$ into the equation for $d$, (\ref{eq:d}). We use the properties of theta functions under translations by a half period to obtain
\be
\frac{\phi_1^\prime(K-\bar{\alpha})}{\phi_1(K-\bar{\alpha})} = \frac{\pi}{2\bar{K}} - \frac{\phi_4^\prime(\bar{\alpha},q)}{\phi_4(\bar{\alpha},q)}.
\ee
Hence, we get
\be
d = \exp\left(-\frac{\pi\bar{\alpha}}{\bar{K}} + 2\frac{\phi_4^\prime(\bar{\alpha},q)}{\phi_4(\bar{\alpha},q)}\,K\right).
\ee
For $q\to0$ the term that multiplies $K$ in the exponential is $O(q)$, while $K$ is $O(\log q)$, and therefore $d$ does not vanish as $q\to0$:
\be
d=-\e^{-2\bar{\alpha}_0}+O(q\log q).
\ee

For $w_\pm\big(x,K-\bar{\alpha}\big)$ we use the relations
\begin{gather}
\phi_1(K-\bar{\alpha},q) = \iu q^{-1/4}\e^{\frac{\pi\bar{\alpha}}{2\bar{K}}} \phi_4(\bar{\alpha},q), \\[2pt]
\phi_2(x\pm K\mp\bar{\alpha},q) = q^{-1/4}\e^{\frac{\pi(x\mp\bar{\alpha})}{2\bar{K}}} \phi_3(x\mp\bar{\alpha},q),
\end{gather}
obtained from the behaviour of theta functions under translations of half period. Then we get that, for $\beta=1$
\be
\begin{split}
w_\pm\big(x,K-\bar{\alpha}\big) & = \pm
\frac{\phi_1^\prime(0,q)}{\phi_4(\bar{\alpha},q)}
\frac{\phi_3(x\pm\bar{\alpha},q)}{\phi_2(x,q)} \,\times \\[2pt]
&\exp\left(\pm\frac{\phi_4^\prime(\bar{\alpha},q)}{\phi_4(\bar{\alpha},q)}\,x\right).
\end{split}
\ee
For $q\to0$, the above expression gives
\be
w_\pm\big(x,K-\bar{\alpha}\big)=(1+\tanh x)\,\e^{-x} + O(q).
\ee
Hence, we obtain
\be
w(x,\beta=1)=C(1+\tanh x)\,\e^{-x} + O(q\log q), \label{eq:wbeta1}
\ee
where $C=1+\e ^{-2\bar{\alpha}_0} = 6-2\sqrt{6}>0$.

\bibliographystyle{unsrt}
\bibliography{references}

 \end{document}